\documentclass[twocolumn,amsmath,amssymb,nofootinbib,prb,floatfix]{revtex4}

\usepackage{graphicx}
\usepackage{dcolumn} 
\usepackage{bm, bbm}      
\usepackage{curves}
\usepackage{epic}
\usepackage{epsf, times}
\usepackage{subfigure}

\begin{document}


\title{ 
Quantum three-coloring dimer model and
the disruptive effect of quantum glassiness 
on its line of critical points 
      }

\author{
Claudio Castelnovo,$^1$
Claudio Chamon,$^1$
Christopher Mudry,$^2$
and
Pierre Pujol,$^3$
       }
\affiliation{
$^1$ Physics Department, Boston University, Boston, MA 02215, USA\\
$^2$
Paul Scherrer Institut, CH-5232 Villigen PSI, Switzerland\\
$^3$ Laboratoire de Physique, Groupe de Physique Th{\'e}orique de
l'{\'E}cole Normale Sup{\'e}rieure, Lyon, France
            }

%
%

\date{\today} 

\begin{abstract}
  We construct a quantum extension of the (classical) three-coloring model
  introduced by Baxter [J.\ Math.\ Phys.\ \textbf{11}, 784 (1970)] for which
  the ground state can be computed exactly along a continuous line of
  Rokhsar-Kivelson solvable points. The quantum model, which admits a local
  spin representation, displays at least three different phases; an
  antiferromagnetic phase, a line of quantum critical points, and a
  ferromagnetic phase. 
  We argue that, in the ferromagnetic phase, the system cannot
  reach dynamically the quantum ground state when coupled to a
  bath through local interactions, and thus lingers in a state of quantum
  glassiness.
\end{abstract}

\maketitle

\section{Introduction}
\label{sec: Introduction}

It has long been known that strongly correlated electronic systems at zero
temperature can be parametrically close to fixed points with
macroscopically degenerate ground state manifolds. 
This is so for a two-dimensional degenerate electron
gas in the quantum Hall regime for which the largest energy scales are the
cyclotron and Zeeman splittings, and for Mott insulators when the Coulomb
energy in the form of an on-site repulsive Hubbard $U$ dominates all other
energy scales.~\cite{Yoshioka02,Wen04} Similarly, the zero temperature phase
diagram of a class of quantum dimer models and geometrically frustrated
quantum systems are known to be influenced by fixed points with massively
degenerate ground state manifolds.~\cite{q_dimer} Whereas most of the efforts
have been directed toward exploring the nature of the collective excitations
present in the different phases and at their boundaries, relatively little
attention has been paid to nonequilibrium phenomena induced by the coupling
to a bath. The goal of this paper is to illustrate how quantum glassiness
can emerge in a two-dimensional homogeneous quantum system \textit{whose
  local degrees of freedom are subjected to a strong constraint}, when the
system is coupled to a bath. In the process, we also show how to obtain a
whole line of solvable Rokhsar-Kivelson (RK) points in a hard-constrained
quantum system.

More specifically, we are going to construct a quantum version of the 
(classical) three-coloring model introduced by Baxter 
in Ref.~\onlinecite{Baxter70}
which we shall call the quantum three-coloring dimer model.
We find that:
(i) For any value of a dimensionless coupling $\beta J$ the ground state
follows from taking a superposition of states in a preferred basis of the
Hilbert space.  The dependence on $\beta J$ of the normalization of the
ground state defines the partition function of a generalization of the
classical three-coloring model at the reduced inverse temperature $\beta J$.%
~\cite{Difrancesco94,Cirillo96,Castelnovo04}
(ii) The excitation spectrum above the ground state is identical to the
spectrum of relaxation times of a classical rate equation that depends on a
dimensionless parameter $\alpha$ in addition to $\beta J$.
(iii) The ground state expectation value of any two local operators at
unequal points in space and imaginary time can be interpreted as the classical
correlation function of the classical three-coloring model endowed with
dynamical rules, provided both operators are diagonal in the preferred basis.
This interplay between a quantum model and a classical stochastic model
generalizes a similar connection established by Henley for
the square lattice quantum dimer model.\cite{Henley97}
(iv)
The quantum three-coloring dimer model is \textit{local} in that it
can be interpreted as a Hamiltonian for quantum spins whose interactions have a 
\textit{finite} range as long as the parameter $\alpha$ remains 
nonnegative. 
In the spin representation, the zero-temperature phase diagram has at least 
three phases, an antiferromagnetic (AF) phase when $-\beta J\gg 1$, 
a line of quantum critical points in the 
neighborhood of infinite 
temperature
and a
ferromagnetic (F) phase when $\beta J\gg1$.  
(v) We argue that, in the F phase, the relaxation time needed for a generic
initial state to decay to the ground state through a \textit{local} 
coupling to a bath diverges exponentially with the system size, 
a property that signals quantum glassiness.~\cite{Chamon04}

\section{Definition of the quantum three-coloring model}
\label{sec: Definition of the quantum three-coloring model}

The classical three-coloring model is a model of dimers laid down on the
bonds of a lattice with coordination number 3. More precisely, 
the dimers come in three kinds (or colors), $A$, $B$, and $C$ 
and occupy the bonds $\langle ij\rangle$ 
between nearest-neighboring sites $i$ and $j$ of the honeycomb
lattice according to two rules: 
(1) all bonds are occupied (close-packed constraint); and
(2) all three dimers meeting at a site (vertex) are different
(three-coloring constraint).
The classical phase space is thus the set
$\mathcal{S}$ of all configurations $\mathcal{C}$ obtained by coloring the
bonds of the honeycomb lattice in any of the three colors $A$, $B$, $C$ with
the condition that three bonds meeting at a site have different colors.  Two
possible configurations $\mathcal{C}^{\ }_{AF}$ and $\mathcal{C}^{\ }_{ F}$
are depicted in Fig.\ \ref{fig: AF 3-coloring conf}.

\begin{figure}[!ht]
\centering
\includegraphics[width=0.45\columnwidth]{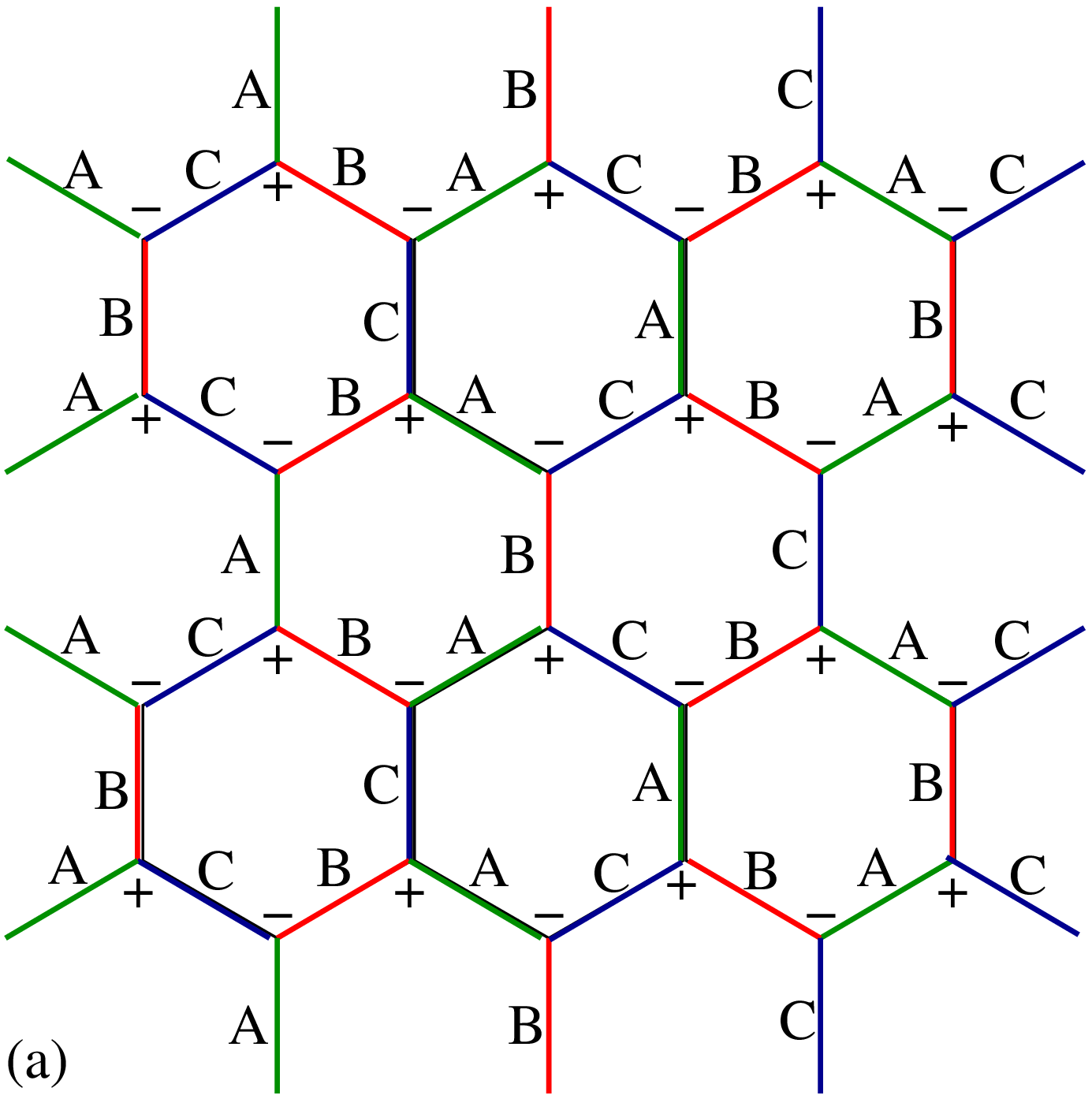}
\quad
\includegraphics[width=0.45\columnwidth]{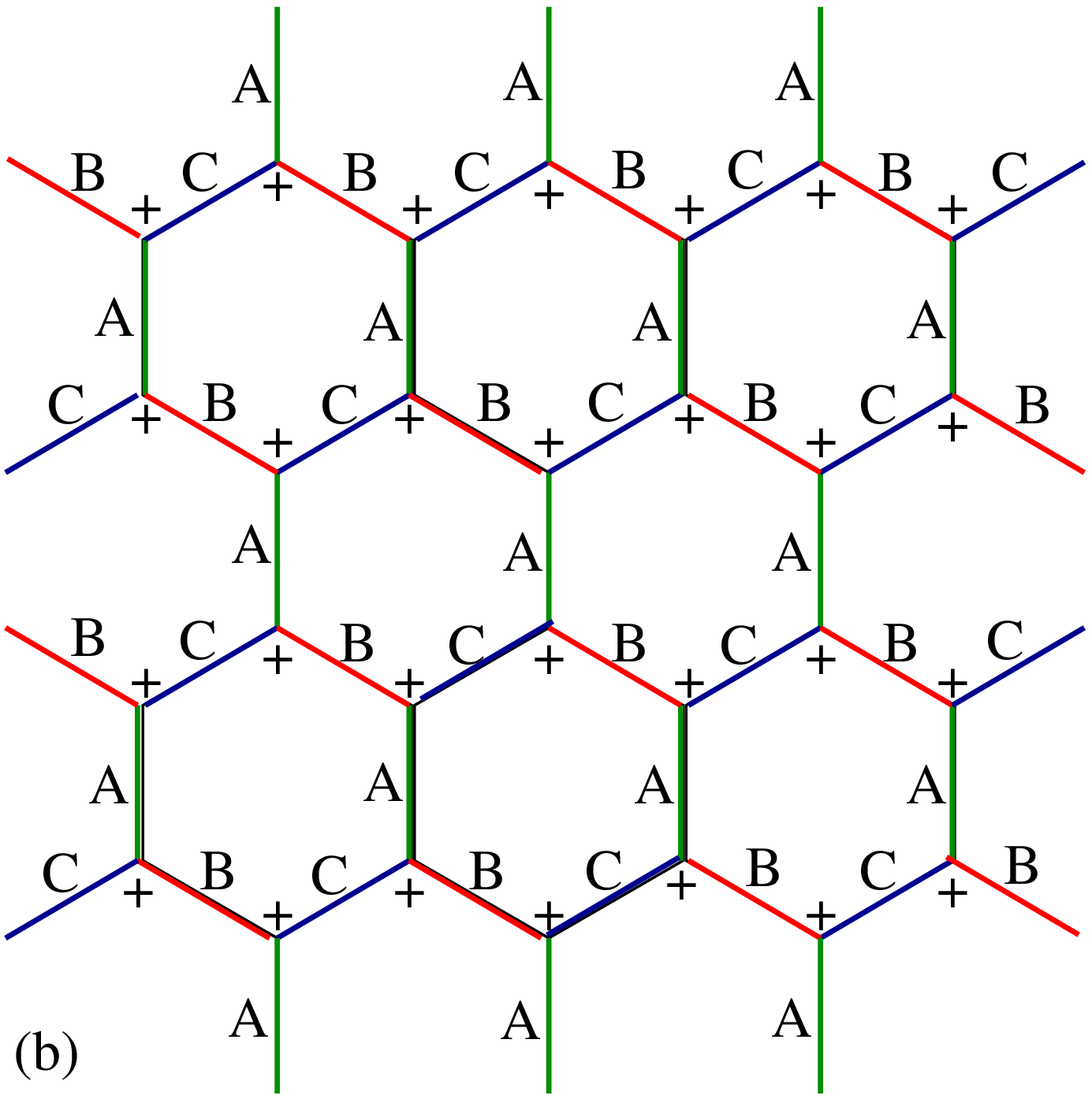}
\caption{
\label{fig: AF 3-coloring  conf}
(Color online) 
Two configurations,
$\mathcal{C}^{\ }_{AF}$ in panel (a)
and 
$\mathcal{C}^{\ }_{ F}$ in panel (b),
from the three-coloring model.
        }
\end{figure}

To any configuration $\mathcal{C}\in\mathcal{S}$ there corresponds a unique
configuration $\{\sigma^{\ }_{i}\}$ of Ising degrees of freedom $\sigma^{\
}_{i}=\pm$, defined on the sites $i$ of the honeycomb lattice, whereby
$\sigma^{\ }_{i}=+(-)$ if the three dimers meeting at $i$ can be obtained
from $ABC$ through an even (odd) permutation when moving clockwise around
$i$.  The configuration $\mathcal{C}^{\ }_{AF}$ ($\mathcal{C}^{\ }_{F}$) in
Fig.~\ref{fig: AF 3-coloring conf} thus corresponds to an AF (F)
configuration of the Ising spins.  Evidently, any global even permutation of
the colors $ABC$ leaves the Ising configuration unchanged. Hence there is a
three-to-one relation between $\mathcal{S}$ and the set of all Ising spin
configurations on the honeycomb lattice such that the magnetization along the
perimeter of any elementary hexagon is either 0 as in 
Fig.~\ref{fig: AF 3-coloring conf}(a) or $\pm6$ as in 
Fig.~\ref{fig: AF 3-coloring conf}(b).\cite{modulo3vsmodulo6} 

Another useful characterization of a configuration
$\mathcal{C}\in\mathcal{S}$ is in terms of \textit{two-color paths}, 
each of which is defined by drawing a line along all bonds colored 
in any given two (say $B$ or $C$) of
the three colors that initiate from a site (the seed)
and by specifying the color pattern (say $BCBC...$ or $CBCB...$) 
along the line.
Upon imposing periodic boundary conditions one can interpret a 
configuration $\mathcal{C}\in\mathcal{S}$ as a
set of two-color closed paths (where the same two colors have been chosen 
for all the paths in the lattice) 
that are nonintersecting and close-packed. 
We will refer to such objects making up a configuration
$\mathcal{C}\in\mathcal{S}$ as \textit{loops}.
For example, configuration $\mathcal{C}^{\ }_{AF}$ ($\mathcal{C}^{\ }_{F}$) in
Fig.~\ref{fig: AF 3-coloring conf} can be viewed as being made solely of
loops of maximal (minimal) 
nonnegative curvature. 

\begin{figure}[!ht]
\centering
\includegraphics[width=0.98\columnwidth]{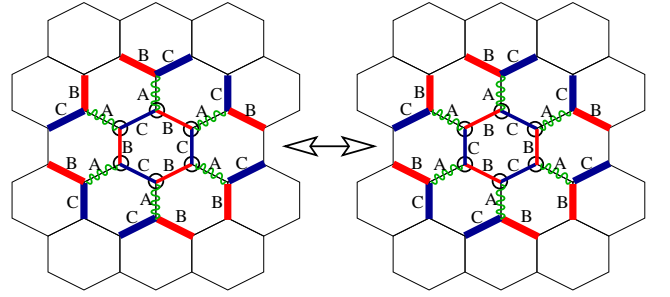}
\caption{
\label{fig: color-swap}
(Color online) 
Two examples of decorated loops of the types $BCBC...$ and $CBCB...$
in the three-coloring model are shown. They are related
by the color-swap defined in the text.
Notice the information encoded in both loops,
(1) the backbone structure (circled lattice sites) 
    that is specified by all the sites 
    visited by the sequence of
    bonds of color $B$ or $C$ emanating from a seed, 
(2) the two-color pattern ($BCBC...$ and $CBCB...$) along each loop, and 
(3) the colors and locations of all the dangling bonds (thick lines). 
The connecting bonds that link 
the backbone of a decorated loop to its 
dangling bonds are represented by wavy lines for clarity.
        }
\end{figure}

We will denote by $\ell$ a {\it decorated loop}, 
i.e., an object made of all $2L^{\ }_{\ell}$ independent colored bonds 
emanating from the $L^{\ }_{\ell}$ sites through which a
loop passes, and by the set of its colored dangling bonds, as defined 
hereafter (see also Fig.~\ref{fig: color-swap}). 
For any $\ell$, the $L^{\ }_{\ell}$ bonds that alternate between
the colors $B$ and $C$, say, form the backbone of $\ell$, while the remaining
$L^{\ }_{\ell}$ bonds of color $A$ that stick out of the backbone 
connect the loop to its dangling bonds, i.e., the two bonds that depart 
from the free end of each $A$ colored bond. 
The definition of a decorated loop therefore encodes the information about 
the colors of all the dangling bonds of the loop. 
The reasons for such a choice will become evident as we proceed with the 
formulation of our quantum model [see Eq.(\ref{eq: def varepsilon}) and 
thereafter].
The set of all possible decorated loops is $\mathcal{L}$. 
Configuration $\mathcal{C}^{\ }_{AF}$ in
Fig.\ \ref{fig: AF 3-coloring conf} is made exclusively of objects $\ell$
with six connecting bonds (i.e., the bonds joining the loop backbone to 
the dangling bonds) of color $A$, $B$, and $C$, respectively.
Configuration $\mathcal{C}^{\ }_{ F}$ in Fig.\ \ref{fig: AF 3-coloring conf}
is made exclusively of objects $\ell$ with $L$ connecting bonds of color $A$,
$B$, and $C$, respectively, where $L$ is the linear size of the system.

By specifying the colors and the locations of both the connecting and the 
dangling bonds, one identifies two decorated loops $\ell$ and $\bar\ell$
related by the \textit{color-swap} operation $B\leftrightarrow C$ along the 
$L^{\ }_{\ell}$ bonds making up the loop backbone. 
This color swap relating two \textit{different} configurations
$\mathcal{C} $ and $\overline{\mathcal{C}}$ in $\mathcal{S}$
is illustrated in Fig.~\ref{fig: color-swap}.
Given a decorated loop $\ell$ and a configuration $\mathcal{C}$, it is always
possible to decide if $\ell$ belongs to $\mathcal{C}$, i.e., either
$\ell\in\mathcal{C}$ or $\ell\notin\mathcal{C}$. By construction: (1) If
$\ell\in\mathcal{C}$ then $\bar\ell\notin\mathcal{C}$.  (2) If
$\ell\in\mathcal{C}$ then there exists a unique configuration
$\overline{\mathcal{C}}$ (to which $\bar\ell$ belongs) obtained by the color
swap operation along the backbone of $\ell$ in $\mathcal{C}$.

To define the quantum three-coloring dimer model we need a Hilbert space and a
Hamiltonian. The Hilbert space is the span of all orthonormal
states $|\mathcal{C}\rangle$ labeled by the classical configurations
$\mathcal{C}\in\mathcal{S}$. The Hamiltonian, which we show below to be of the
RK type, takes the form
\begin{subequations}
\label{eq: def RK hamiltonian}
\begin{eqnarray}
&&
\widehat{H}^{\ }_{RK}:=
\sum_{\ell\in\mathcal{L}}
w^{\ }_{\ell}
\widehat{Q}^{\ }_{\ell},
\label{eq: def RK hamiltonian a}
\\
&&
\widehat{Q}^{\ }_{\ell}:=
\widehat{V}^{\ }_{\ell}
-
\widehat{T}^{\ }_{\ell,\bar\ell},
\label{eq: def RK hamiltonian b}
\end{eqnarray}
where the potential and kinetic energy operators are
\begin{eqnarray}
&&
\widehat{V}^{\ }_{\ell}:=
e^{
-\beta J\varepsilon^{\ }_{\ell}/2
  }
\widehat{P}^{\ }_{\ell}\equiv
e^{
-\beta J\varepsilon^{\ }_{\ell}/2
  }
\sum_{\mathcal{C}\in\mathcal{S}}
c^{\ }_{\ell}
|\mathcal{C}\rangle
\langle\mathcal{C}|,
\label{eq: def RK hamiltonian c}
\\
&&
\widehat{T}^{\ }_{\ell,\bar\ell}:=\sum_{\mathcal{C}\in\mathcal{S}}
c^{\ }_{\ell}
|\overline{\mathcal{C}}\rangle
\langle\mathcal{C}|,
\label{eq: def RK hamiltonian d}
\end{eqnarray}
respectively.
The dimensionless coupling $\beta J$ is real valued and
$c^{\ }_{\ell}=1$ if $\ell\in\mathcal{C}$, while it vanishes otherwise.
The dimensionless energy 
\begin{eqnarray}
E^{\ }_{\mathcal{C}}:=
-\sum_{\langle ij\rangle} \sigma^{\ }_{i}\sigma^{\ }_{j}
\label{eq: def EC}
\end{eqnarray}
is the classical energy of the Ising spin configuration 
$\{\sigma^{\ }_{i}\}$
associated to
$\mathcal{C}$
in units of $|J|$. Finally, the dimensionless energy 
\begin{eqnarray}
\varepsilon^{\ }_{\ell}=
E^{\ }_{\overline{\mathcal{C}}}
-
E^{\ }_{\mathcal{C}}
\label{eq: def varepsilon}
\end{eqnarray}
\end{subequations}
is uniquely defined for any $\mathcal{C}\in\mathcal{S}$ that contains $\ell$. 
Notice that the definition of a decorated loop was precisely chosen so that 
the energy difference 
$E^{\ }_{\overline{\mathcal{C}}}
-
E^{\ }_{\mathcal{C}}$ 
depends only on the information contained in $\ell$, and not on further 
details of $\mathcal{C}$ and $\overline{\mathcal{C}}$. 

Whatever the choice made for 
the energy scales $\{w^{\ }_{\ell}\geq0\}$,
the wave function
\begin{subequations}
\begin{eqnarray}
&&
|\Psi^{\ }_{RK}\rangle=
\sum_{\mathcal{C}\in\mathcal{S}}
\exp\left(-\beta J E^{\ }_{\mathcal{C}}/2\right)
|\mathcal{C}\rangle
\label{eq: def RK ground state}
\end{eqnarray}
is the ground state of $\widehat{H}^{\ }_{RK}$,
as follows from the facts that: 
(1) $\widehat{Q}^{2 }_{\ell}$ is proportional to $\widehat{Q}^{\ }_{\ell}$, 
and 
(2) $|\Psi^{\ }_{RK}\rangle$ is annihilated by $\widehat{Q}^{\ }_{\ell}$ 
($\forall \, \ell\in\mathcal{L}$). 
Remarkably, the normalization of 
$\langle\Psi^{\ }_{RK}|\Psi^{\ }_{RK}\rangle$ is nothing
but the partition function 
\begin{eqnarray}
Z:=\sum_{\mathcal{C}\in\mathcal{S}}
e^{-\beta J E^{\ }_{\mathcal{C}}}
\label{eq: 3-color Z}
\end{eqnarray}
\end{subequations}
of the generalized three-coloring model
(see Refs.~\onlinecite{Difrancesco94,Cirillo96,Castelnovo04}) 
with the configuration
energy (\ref{eq: def EC}), i.e., up to a multiplicative factor of 3, it is
the partition function of the nearest-neighbor Ising model on the honeycomb
lattice with the \textit{local constraint} that the magnetization on any
elementary hexagon is 0 or $\pm6$.\cite{modulo3vsmodulo6}
The zero-temperature phase
diagram of the quantum three-coloring model along the axis $\beta
J\in\mathbb{R}$ \textit{contains} the phase diagram of the generalized
classical three-coloring model along the reduced temperature axis $\beta
J\in\mathbb{R}$.

\section{Phase diagram of the classical three-coloring model}
\label{sec: Phase diagram of the classical three-coloring model}

The partition function (\ref{eq: 3-color Z}) was evaluated by Baxter at
infinite temperature.~\cite{Baxter70} Through a height representation of the
three-coloring model, it was shown in Ref.~\onlinecite{Kondev96} that at
infinite temperature the probability that two points of the honeycomb lattice
separated by a distance $r$ belong to the same loop decays with $r$
algebraically and that this decay is captured by the SU(3)$^{\ }_{1}$
Wess-Zumino-Witten (WZW) theory.  The finite temperature phase diagram was
studied numerically in Ref.~\onlinecite{Difrancesco94} with a transfer matrix 
approach, in Ref.~\onlinecite{Cirillo96} with a cluster variational method 
(CVM), and in Ref.~\onlinecite{Castelnovo04} using Monte Carlo simulations, 
the CVM, as well as perturbative renormalization-group (RG) techniques 
for infinitesimal values of the 
coupling $\beta J$. The classical 
configuration that minimizes the energy $E^{\ }_{\mathcal{C}}$ is the
antiferromagnatic $\mathcal{C}^{\ }_{AF}$ when $J<0$ and the ferromagnetic
$\mathcal{C}^{\ }_{ F}$ when $J>0$ (see Fig.~\ref{fig: AF 3-coloring conf}).
Thermal fluctuations are expected to destroy both long-range ordered phases
for a sufficiently large temperature.  
In view of the \textit{local constraint} of the three-coloring model, 
the transition may be different in nature from the one of the 
unconstrained model. Numerical simulations are
consistent with the termination of the F phase in a first order phase
transition at the reduced temperature $(\beta J)^{\ }_{F}$.%
~\cite{Castelnovo04} 
On the other hand, the SU(3)$^{\ }_{1}$ 
WZW critical point at infinite temperature seems to be the 
end point of a line of critical points when 
$0<\beta J\leq
(\beta J)^{\ }_{\mathrm{cr}}$.  Whether $(\beta J)^{\ }_{\mathrm{cr}}=(\beta
J)^{\ }_{F}$ or there exists intermediary phases~\cite{Fradkin04,Vishwanath04} 
in the range $(\beta J)^{\ }_{\mathrm{cr}}<\beta J< (\beta J)^{\ }_{F}$ 
is not known at present. The phase diagram is not symmetric under $J\to-J$ as 
the \textit{local constraint} favors the AF over the F state from an entropic
point of view.  The CVM predicts a critical end point $(\beta J)^{\ }_{AF}<0$
to the AF phase followed by a paramagnetic phase, possibly of the short-range
resonating-valence-bond (RVB) type, that terminates in the critical point at
infinite temperature.  The conjectured phase diagram of the partition
function (\ref{eq: 3-color Z}) is sketched in Fig.~\ref{fig: phase
  diagram}.

\begin{figure}[!ht]
\centering
\includegraphics[width=0.95\columnwidth]{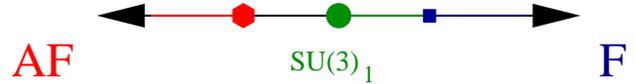}
\caption{\label{fig: phase diagram}
(Color online) 
Conjectured phase diagram of
the classical partition function (\ref{eq: 3-color Z})
along the reduced temperature axis $\beta J$.
The right end point of the AF phase is a hexagon.
The left and right end points of the critical phase are a 
circle and a square, respectively. The circle corresponds
to $\beta J=0$ and realizes the SU(3)$^{\ }_{1}$ WZW critical theory.
        }
\end{figure}

\section{
Excitation spectrum, locality, and emerging topological quantum numbers
        }
\label{sec: Excitation spectrum, locality, and emerging topological}

Although the ground state of $\widehat{H}^{\ }_{RK}$ does not depend on the
choice of the couplings $w^{\ }_{\ell}$, the excitation spectrum does. We are
now going to show that there exists a one-to-one correspondence between the
excitation energies of $\widehat{H}^{\ }_{RK}$ and the relaxation modes of
the three-coloring model endowed with the dynamics borrowed from a uniquely
defined transition matrix $(W^{\ }_{\mathcal{C}\mathcal{C}'})$. We
begin by defining the matrix elements
\begin{subequations}
\begin{eqnarray}
W^{\ }_{\mathcal{C}\mathcal{C}^{\prime}}&:=&
-
e^{-
\beta J
\left(E^{\ }_{\mathcal{C}}-E^{\ }_{\mathcal{C}^{\prime}}\right)/2
  }
\langle\mathcal{C}|
\widehat H^{\ }_{RK}
|\mathcal{C}^{\prime}\rangle.
\end{eqnarray}
Although $(W^{\ }_{\mathcal{C}\mathcal{C}'})$
is not symmetric in general, it shares the same set of eigenvalues
as $\widehat H^{\ }_{RK}$. 
Furthermore it satisfies by construction
the detailed balance condition 
\begin{eqnarray}
&&
W^{\ }_{\mathcal{C}\mathcal{C}^{\prime}} 
e^{-\beta J E^{\ }_{\mathcal{C}^{\prime}}}=
W^{\ }_{\mathcal{C}^{\prime}\mathcal{C}} 
e^{-\beta J E^{\ }_{\mathcal{C}}}
\label{eq: detailed balance W} 
\end{eqnarray}
as well as the condition
\begin{eqnarray}
0=
\sum_{\mathcal{C}^{\prime}\in\mathcal{S}}
W^{\ }_{\mathcal{C}^{\prime}\mathcal{C}}.
\label{eq: sum over rows =0}
\end{eqnarray}
\end{subequations}
Equation (\ref{eq: sum over rows =0})
allows one to interpret the first order
linear differential equation in imaginary time $\tau$
\begin{eqnarray}
\dot{p}^{\ }_{\mathcal{C}}(\tau)=
\sum_{\mathcal{C}^{\prime}\in\mathcal{S}}
W^{\ }_{\mathcal{C}\mathcal{C}^{\prime}}
p^{\ }_{\mathcal{C}^{\prime}}(\tau)
\label{eq: master equations as a matrix equation}
\end{eqnarray}
as a master equation for the probabilities 
$\{p^{\ }_{\mathcal{C}}(\tau)\}$
whose stationary solution is
\begin{equation}
p^{(0)}_{\mathcal{C}}=Z^{-1}\exp(-\beta J E^{\ }_{\mathcal{C}})
\end{equation}
by Eq.\ (\ref{eq: detailed balance W}).
Furthermore,
the properly normalized ground state expectation value
$\widehat{G}^{\ }_{\widehat{B}\widehat{A}}(\tau)$
of the product 
$\widehat{B}(\tau)\times\widehat{A}(0)$
at unequal imaginary times
of any two operators 
$\widehat{A}$
and
$\widehat{B}$
diagonal in the basis $|\mathcal{C}\rangle$
can be interpreted as a classical dynamical correlation function 
at unequal times: 
\begin{eqnarray}
\widehat{G}^{\ }_{\widehat{B}\widehat{A}}(\tau)&=&
\sum_{\mathcal{C}^{\prime}}
\sum_{\mathcal{C}}
B^{\ }_{\mathcal{C}^{\prime}}
p^{\ }_{\mathcal{C^{\prime}|\mathcal{C}}}(\tau)
A^{\ }_{\mathcal{C}}
p^{(0)}_{\mathcal{C}}
\equiv
G^{\ }_{BA}(\tau).
\label{eq: classical correlation}
\end{eqnarray}
The conditional probability 
$p^{\ }_{\mathcal{C^{\prime}|\mathcal{C}}}(\tau)$
is the solution to the master equation 
(\ref{eq: master equations as a matrix equation})
with the initial condition 
$
p^{\ }_{\mathcal{C}^{\prime}|\mathcal{C}}(\tau=0)=
\delta^{\ }_{\mathcal{C}^{\prime}\mathcal{C}}
$
while $B^{\ }_{\mathcal{C}^{\prime}}$
and   $A^{\ }_{\mathcal{C}}$
are the matrix elements of $\widehat{B}$ and $\widehat{A}$, respectively.
Equation (\ref{eq: classical correlation}),
relating classical stochastics of the three-coloring model to the 
quantum dynamics in imaginary time of the quantum three-coloring dimer model, 
is an example of a generic property of quantum dimer models at the 
so-called RK point as noted by Henley.~\cite{Henley97}

We can now motivate our choice 
for the couplings $w^{\ }_{\ell}$, 
\begin{eqnarray}
w^{\ }_{\ell}=
w^{\ }_{0}
f^{\ }_{\ell}
\exp\left[-\alpha \left(L^{\ }_{\ell}-6\right)\right].
\label{eq: def w of ell}
\end{eqnarray}
Here, the energy scale $w^{\ }_{0}$ is of order 1.
The number $0< f^{\ }_{\ell}\leq 1$
can always be chosen such that the classical stochastics in time encoded by
Eq.\ (\ref{eq: master equations as a matrix equation}) 
is the Glauber or Metropolis dynamics.~\cite{Castelnovo05}
We do not believe this choice to be of consequence for spectral
properties beyond the mean level spacing. The choice for the value
of the dimensionless parameter $\alpha$ can affect spectral
properties in a crucial way, however. 
Any $\alpha>0$ implements (defines), in the thermodynamic limit,
the condition of locality. A corollary to $\alpha>0$ is the emergence of 
well-defined topological sectors in the thermodynamic limit.

The quantum three-coloring Hamiltonian 
(\ref{eq: def RK hamiltonian})
where the coupling constants $w^{\ }_{\ell}$ are defined in
Eq.~(\ref{eq: def w of ell}) with $\alpha>0$
is local in very much the same way as the effective
quantum spin-1/2 model obtained
from the expansion of the Hubbard model at half-filling, and in the limit
of an on-site repulsion $U$ much larger than the band width $t$.
The potential energy~(\ref{eq: def RK hamiltonian c})
associated with a decorated loop $\ell$ with perimeter $L^{\ }_{\ell}$
generates 2, 3, ..., $L^{\ }_{\ell}$-body Ising spin interactions
for spins that can be as far as $L^{\ }_{\ell}$ lattice spacings apart
very much in the same way as ring exchange interactions along a closed path
of length $L^{\ }_{\ell}$ 
are induced to $L^{\ }_{\ell}$-th order in $t/U$ for the Hubbard model
at half-filling. In the case of both the quantum three-coloring and the
effective Heisenberg model, these $L^{\ }_{\ell}$-body interactions 
are associated to characteristic energy scales that are exponentially 
suppressed with $L^{\ }_{\ell}$; namely, 
$w^{\ }_{\ell}$ and $(t/U)^{L^{\ }_{\ell}}\times t$, respectively. 
The same argument applies to the kinetic 
energy~(\ref{eq: def RK hamiltonian d}). 
In fact, we shall show shortly that here the counterpart to 
the Hubbard model is the constrained transverse field Ising on the 
honeycomb lattice. 

\begin{figure}[!ht]
\centering
\includegraphics[width=0.45\columnwidth]{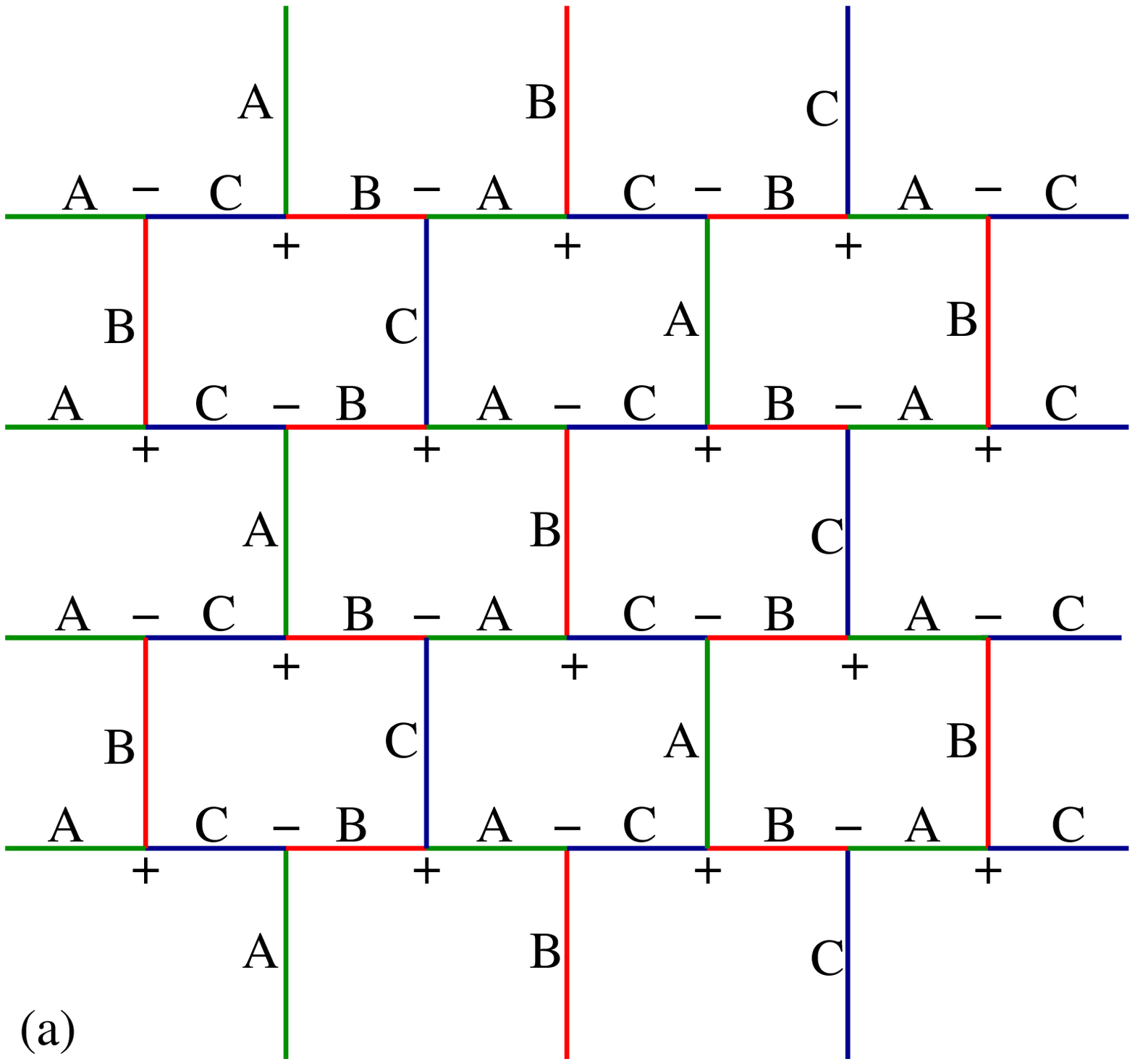}
\quad
\includegraphics[width=0.45\columnwidth]{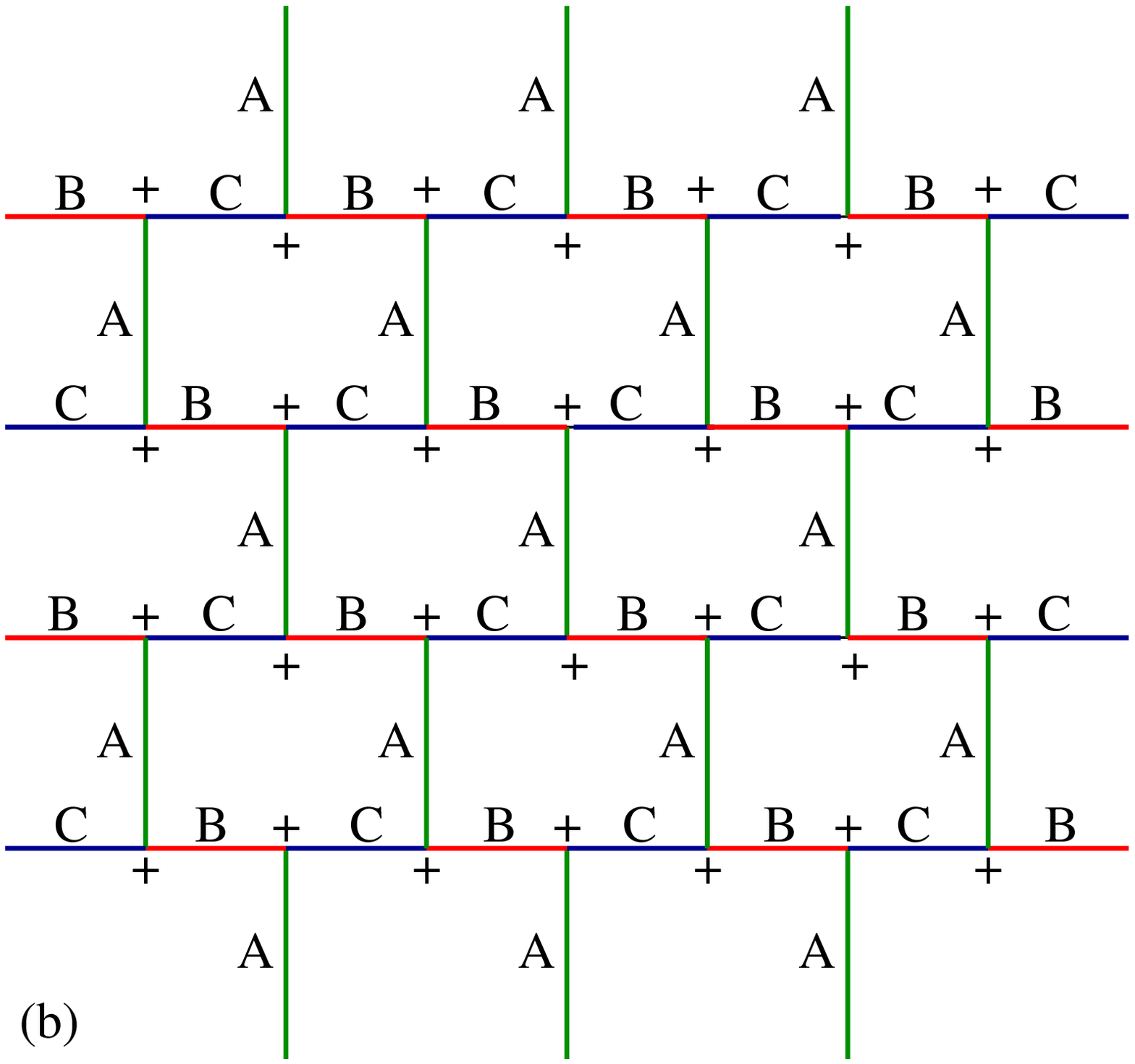}
\caption{
\label{fig: 3-coloring  conf brick wall}
(Color online) 
Two configurations
$\mathcal{C}^{\ }_{AF}$ in panel (a)
and 
$\mathcal{C}^{\ }_{ F}$ in panel (b)
in the brick wall representation.
In panel (a)
$N^{\ }_{A}=N^{\ }_{B}=N^{\ }_{C}$.
In panel (b)
$N^{\ }_{A}$ is maximum with 
$N^{\ }_{B}=N^{\ }_{C}=0$ for $\mathcal{C}^{\ }_{F}$.
        }
\end{figure}

The concept of topological sectors is rooted in the following conservation
laws characterizing any $\mathcal{C}\in\mathcal{S}$.~\cite{Baxter70}
Choose the brick wall representation of the three-coloring model as depicted 
in Fig.\ \ref{fig: 3-coloring conf brick wall}. Draw a line between any two
consecutive rows and count the number of intersections $N^{\ }_{A,B,C}$ with
vertical bonds of color $A$, $B$, and $C$, respectively. This triplet of
numbers is independent of the choice of the two consecutive rows and
characterizes globally each configuration $\mathcal{C}\in\mathcal{S}$.
Furthermore, for any $\ell\in\mathcal{L}$ and for any
$\mathcal{C}\in\mathcal{S}$ such that $\ell\in\mathcal{C}$, $N^{\
}_{A,B,C}=\overline{N}^{\ }_{A,B,C}$ whereby $N^{\ }_{A,B,C}$
($\overline{N}^{\ }_{A,B,C}$) characterizes globally $\mathcal{C}$
($\overline{\mathcal{C}}$), as long as $\ell$ is nonwinding. The numbers
$N^{\ }_{A,B,C}$ are elevated to the status of conserved quantum numbers
under the quantum dynamics defined by $\widehat{H}^{\ }_{RK}$, provided $w^{\
}_{\ell}=0$ for all winding loops. In particular, these topological
conservation laws emerge in the thermodynamic limit when $\alpha>0$.  It is
possible that $0\leq \beta J\leq (\beta J)^{\ }_{\mathrm{cr}}$ realizes
a line of quantum Lifshitz critical points when $\alpha>0$,~\cite{Ardonne04}
with a dynamical exponent $z=2$; this is in principle testable
numerically.

\section{
Quantum three-coloring model and the constrained quantum Ising model
in a transverse field
        }

The RK three-coloring Hamiltonian
(\ref{eq: def RK hamiltonian a})
is a special case of the Hamiltonian
\begin{eqnarray}
\widehat{H}^{\ }_{3-c}:=
\sum_{\ell\in\mathcal{L}}
w^{\ }_{\ell}
\left(
v^{\ }_{\ell}
\widehat{P}^{\ }_{\ell}
-
t^{\ }_{\ell}
\widehat{T}^{\ }_{\ell,\bar\ell}
\right)
\label{eq: generalized 3-c Hamiltonian}
\end{eqnarray}
where the dimensionless couplings $v^{\ }_{\ell}$ and $t^{\ }_{\ell}$ 
are real valued. The choice of the couplings 
$v^{\ }_{\ell}=-\varepsilon^{\ }_{\ell}J/(4w^{\ }_{\ell})$
and
$t^{\ }_{\ell}=1$
implies that: 
(1) 
the matrix element
$$
-(J/4)
\sum_{\ell\in\mathcal{L}}
\varepsilon^{\ }_{\ell}
\langle\mathcal{C}|
\widehat{P}^{\ }_{\ell}
|\mathcal{C}\rangle
$$
is nothing but the classical Ising energy
$$
-J\sum_{\langle ij\rangle}\sigma^{\rm z}_{i}\sigma^{\rm z}_{j}
$$
of the nearest-neighbor Ising model on the honeycomb lattice
subjected to the \textit{local constraint on any elementary hexagon}%
\cite{modulo3vsmodulo6}
\begin{equation}
\cos\left(2\pi\sum_{i\in\mathrm{hex}}\sigma^{\rm z}_{i}/3\right)=1
\end{equation}
while
(2) 
each term
$
w^{\ }_{\ell}
\widehat{T}^{\ }_{\ell,\bar\ell}
$
is induced order by order in degenerate perturbation theory by
the \textit{small transverse field} 
$\Gamma\sum_{i}\sigma^{\rm x}_{i}$.
Here, it is the local constraint that is responsible for the 
macroscopic quasidegeneracy of the classical limit $\Gamma\to0$
as opposed to the geometrical frustration in the Ising models treated
in Ref.~\onlinecite{Moessner00}. This happens as long as 
the energy splitting of order $J$ characteristic of the classical 
configurations that obey the three-coloring constraint is negligible 
compared to the very large characteristic energy $U$ paid by any 
classical configuration that violates this constraint. The energy scale 
associated to the three-coloring constraint plays a role similar to 
that of a large Hubbard on-site repulsion for strongly correlated electrons 
or of the cyclotron energy in the regime of the fractional Hall effect. 
The constrained quantum Ising Hamiltonian in a small transverse field, 
\begin{equation}
\widehat{H}^{\ }_{\textrm{Ising}}:=
\sum_{\ell\in\mathcal{L}}
\left(
-(J/4)
\varepsilon^{\ }_{\ell}
\widehat{P}^{\ }_{\ell}
-
w^{\ }_{\ell}
\widehat{T}^{\ }_{\ell,\bar\ell}
\right),
\label{eq: def H_Ising}
\end{equation}
captures the $\mathbb{Z}^{\ }_{2}$ quantum dynamics 
induced by time-reversal symmetry breaking
in a triangular array of 
Josephson junctions when the superconducting order parameter breaks 
time-reversal invariance.~\cite{Castelnovo04,Moore04}
The quantum three-coloring Hamiltonians 
$\widehat{H}^{\ }_{3-c}$
and $\widehat{H}^{\ }_{RK}$
are also relatives to a model for quantum spin-1 degrees of freedom
defined on the bonds of the honeycomb lattice
introduced by Wen in Ref.~\onlinecite{Wen03}.

We can now refine the notion of locality $(\alpha>0)$ of $\widehat{H}^{\ }_{RK}$. 
The quantum dynamics in real time of $\widehat{H}^{\ }_{RK}$ 
and of the constrained quantum 
Ising model $\widehat{H}^{\ }_{\textrm{Ising}}$
originate from the very same quantum tunneling processes
between classical configurations. 
Their local nature is explicit in
$\widehat{H}^{\ }_{\textrm{Ising}}$ 
and carries over to 
$\widehat{H}^{\ }_{RK}$. 

\section{
Local coupling to a thermal bath, arrested quantum dynamics or
quantum glassiness
        }
\label{sec: Local coupling to a thermal bath, arrested quantum dynamics}

So far we have considered the quantum three-coloring model
(\ref{eq: generalized 3-c Hamiltonian})
or, equivalently, the constrained Ising model in a transverse field
(\ref{eq: def H_Ising}) 
in isolation. 
Energy is then a conserved quantum number and the real-time evolution of
any state in the Hilbert space is a linear superposition of harmonics with
frequencies $\omega^{\ }_{j}=E^{\ }_{j}/\hbar$ where the index
$j$ labels the exact eigenvalues of Hamiltonians 
(\ref{eq: generalized 3-c Hamiltonian})
or
(\ref{eq: def H_Ising}).

In the spirit of Caldeira and Legget,\cite{Caldeira83}
imagine coupling 
$\widehat{H}^{\ }_{\textrm{Ising}}$
($\widehat{H}^{\ }_{RK}$)
\textit{very weakly} ($g^{\ }_{i\lambda^{\ }_{i}}\ll J,w^{\ }_{0}$)
to a bath of harmonic oscillators 
$\{a^{\dag}_{i,\lambda^{\ }_{i}},a^{\ }_{i,\lambda^{\ }_{i}}\}$
through the \textit{local} interactions~\cite{Chamon04} 
\begin{equation}
\label{eq: bath coupling}
\sum_{i\lambda^{\ }_{i}} 
  g^{\ }_{i\lambda^{\ }_{i}} \sigma^{\rm x}_{i}
    \left(
      a^{\ }_{i,\lambda^{\ }_{i}}+ \mathrm{h.c}
    \right). 
\end{equation}
The effect of this phonon bath is to endow the energy eigenvalues of
the strongly correlated system in isolation with finite lifetimes.
These lifetimes can be calculated perturbatively. To put it
differently, the presence of a phonon bath implies 
that we cannot consider a pure state, 
as was the case in deriving Eq.~(\ref{eq: classical correlation}).  
The introduction of a density matrix is now required
to study the thermodynamic behavior of 
$\widehat{H}^{\ }_{\textrm{Ising}}$
($\widehat{H}^{\ }_{RK}$)
and, in this sense, the evolution in real time of 
$\widehat{H}^{\ }_{\textrm{Ising}}$
($\widehat{H}^{\ }_{RK}$)
is better understood in terms of transition or relaxation rates 
between states of
$\widehat{H}^{\ }_{\textrm{Ising}}$
($\widehat{H}^{\ }_{RK}$)
by absorption or emission of phonons. At zero temperature,
any initial state of 
$\widehat{H}^{\ }_{\textrm{Ising}}$
($\widehat{H}^{\ }_{RK}$)
which is different from the ground state of 
$\widehat{H}^{\ }_{\textrm{Ising}}$
($\widehat{H}^{\ }_{RK}$)
can now decay to the ground state through spontaneous emission 
of phonons ($a^{\dag}_{i,\lambda^{\ }_{i}}$), 
i.e., by transferring energy to the thermal bath of harmonic 
oscillators.
The goal of this section is to estimate the relaxation rate of
an initial state of 
$\widehat{H}^{\ }_{\textrm{Ising}}$
($\widehat{H}^{\ }_{RK}$)
into the ground state of
$\widehat{H}^{\ }_{\textrm{Ising}}$
($\widehat{H}^{\ }_{RK}$)
due to the coupling to the bath.
It should be emphasized that these relaxation rates are 
completely unrelated to the imaginary-time dynamics of 
$\widehat{H}^{\ }_{RK}$ 
encoded by 
Eq.~(\ref{eq: classical correlation}). 

Deep in the F phase ($J\gg w^{\ }_{0}$) all the loops are 
straight and winding around the system (see configuration 
$\mathcal{C}_F$ in Fig.~\ref{fig: AF 3-coloring  conf}). 
The relaxation time for an initial state to decay into the F ground 
state is bounded from below by a time of the order of the 
time needed by any state 
$|\Psi^{\ }_{F,\ell\in\mathcal{C}_F}\rangle = 
\widehat{T}^{\ }_{\ell,\bar\ell} |\Psi^{\ }_{F}\rangle$ 
to decay into the F ground state $|\Psi^{\ }_{F}\rangle$ 
through spontaneous emission of phonons.
The matrix element for such a quantum tunneling process is obtained 
from the $L$-th order perturbation expansion of the 
coupling~(\ref{eq: bath coupling}), and it is therefore suppressed by 
an exponential factor of order $(g/U)^L$, where $U$ is now the (large) 
characteristic energy scale for the three-coloring constraint and $g$ 
is the typical coupling to the bath [notice that this process requires 
${\cal O}(L)$ phonons]. 
Hence the relaxation time for an arbitrary initial state to decay 
through spontaneous emission of phonons into the F ground state is 
bounded from below by some positive power of $\exp(L)$.~\cite{Chamon04} 
The same rational applies to $\widehat{H}^{\ }_{RK}$. 

Exponential dependence of the relaxation times on the system size 
is a signature of the breakdown of thermodynamic equilibration 
(arrested quantum dynamics). 
For both $\widehat{H}^{\ }_{RK}$ and 
$\widehat{H}^{\ }_{\textrm{Ising}}$, it occurs in the zero 
temperature phase diagram of Fig.~\ref{fig: phase diagram} 
as soon as the ground and first excited states belong to 
different topological sectors. 
This is a direct consequence of the fact that changing the topological 
sector of a given state 
$\bigotimes^N_{i=1} |\sigma^{\textrm{z}}_{i}\rangle$ 
requires flipping all the values of the $\sigma^{\textrm{z}}_i$ 
(via the corresponding $\sigma^{\textrm{x}}_i$ operators) along winding 
loops that have a length of order $L$. 
In the systems considered here, it so happens that the change in 
topological sector of the ground state occurs only when the system 
orders ferromagnetically, i.e., for sufficiently large $\beta J$. 
In fact, for all $\beta J<(\beta J)^{\ }_{\mathrm{cr}}$ the ground 
and first excited states belong to the topological sector of the AF 
state, i.e., the typical loops that connect the ground 
to the first excited states have a perimeter of order $1$, 
and no arrested quantum dynamics is to be expected. 
For sufficiently large $\beta J$ the ground state belongs to the 
topological sector of the F state while \textit{generic} excited states 
belong to different topological sectors and arrested quantum dynamics 
rules. 

An intriguing question is what is the fate of this arrested quantum 
dynamics when we forbid tunneling processes between different 
topological sectors \textit{before} taking the thermodynamic limit, 
and work exclusively in the topological sector of the AF state. 
We believe that the mechanism for arrested quantum dynamics 
outlined above survives because the ground state for sufficiently 
large $\beta J$ should have a finite overlap 
with a \textit{polycrystalline F state}, 
i.e., a state made of macroscopically large ferromagnetic domains 
in the Ising representation or one made almost exclusively of very 
large loops in the loop representation 
(see Fig.\ 9 of Ref.~\onlinecite{Castelnovo04}).
Indeed, this polycrystalline F state can be obtained from a generic 
excited state of the AF state (characterized exclusively by small loops) 
only by applying $\sigma^{\textrm{x}}_i$ operators along very large 
(though not system-spanning) loops, i.e., after the 
spontaneous emission of a number of phonons of order $L$. 

Dynamical arrest is one of the characteristics of glassy systems. Classical
glass formers, for example, fall out of equilibrium when the temperature is
lowered and the relaxation times become of the order of the experimentally
accessible time scales. 
In classical systems the fast decrease of the relaxational rates
with decreasing temperature is associated to the increase in the times 
needed to surpass energy barriers by thermal activation. In quantum systems, 
tunneling under high (but narrow) activation barriers remains an open 
channel for equilibration down to zero temperature. Such quantum tunneling 
processes are the working principle behind the notion of 
\textit{quantum annealing},~\cite{Kadowaki-Nishimori98} 
which was demonstrated experimentally in the disordered Ising magnet
LiHo$_{0.56}$Y$_{0.44}$F$_{4}$.~\cite{Aeppli99} 
However, quantum annealing is rendered inoperative 
if the tunnel barriers are \textit{wide}, and classical
thermal activation may be a more efficient thermalization process even at
rather low temperatures. In this case, quantum mechanics does not prevent
systems from remaining in a state of glassiness at low temperatures, 
very much as their classical counterparts can do in the presence
of strong frustration induced by geometry or disorder. 

What makes such a state of \textit{quantum glassiness} remarkable is that it 
can occur in a system devoid of geometric frustration or
disorder, solely as a result of strong constraints.
At zero temperature, as thermal activation over energy barriers is shut down,
the only mechanism for relaxation is quantum tunneling. In the quantum 
three-coloring model, the effective tunneling barrier widths are related to 
the sizes of the loops that are color-flipped. As the system tries to decay 
into \textit{polycrystalline $F$ states} with larger and larger grain sizes, 
tunneling of larger and larger objects is needed, 
and hence the relaxation rates grow exponentially with the size of the loops 
as long as the quantum dynamics is of a local nature. 
The extreme case occurs when a generic out-of-equilibrium state attempts to 
decay into the $F$ state, a process that requires the exchange of the colors 
along loops with a perimeter of the order of the linear size of the system. 

We have focused on dynamical arrest as a signature of glassy behavior. One
may wonder whether a definition of a glass requires a thermodynamical
transition into a glassy phase. The answer is negative, as many examples have
been constructed of classical systems with trivial thermodynamics that display
relaxation rates characteristic of glasses of the 
strong and fragile types.~\cite{Ritort-Sollich03}
Reference~\onlinecite{Chamon04} provides one example of a quantum system with 
no thermodynamic transition but characterized by relaxation rates of 
a strong glass type; 
this system defies quantum annealing 
because the tunneling barriers become increasingly wide
as the system effective temperature is lowered, and hence it remains in a
state of quantum glassiness. The quantum three-coloring model is a second 
example thereof. In both cases, quantum glassiness emerges from the local 
nature of the Hamiltonian combined with the presence of strong constraints 
on the quantum dynamics.

\section{Conclusions}
\label{sec: Conclusions}

In this paper we have constructed a specific quantum extension 
$\widehat{H}^{\ }_{RK}$
to the three-coloring model of Baxter. Our strategy is very general as it applies
to any space $\mathcal{S}$
of classical configurations $\mathcal{C}$
and to any chosen subset $\mathcal{L}\in\mathcal{S}\times\mathcal{S}$ 
that defines which pairs of classical configurations are related 
by quantum tunneling. The strategy is specific in that it demands choosing
(1) a potential term
$\widehat{V}(\mathcal{C})$, $\forall\mathcal{C}\in\mathcal{S}$
and
(2) a kinetic term $\widehat{T}(\overline{\mathcal{C}},\mathcal{C})$, 
$\forall(\overline{\mathcal{C}},\mathcal{C})\in\mathcal{L}$.
Our choice for $\widehat{H}^{\ }_{RK}$ was simple in that
the ground state is known for all values of two dimensionless
couplings $\alpha\geq0$ and $\beta J\in\mathbb{R}$.
The zero-temperature phase diagram of $\widehat{H}^{\ }_{RK}$
as a function of $\beta J\in\mathbb{R}$
is intricate. Long-ranged ordered phases in the local Ising variables
at $\beta J\ll-1$ and $\beta J\gg+1$
are separated by qualitatively different
paramagnetic phases of which some are critical with respect to equal-time
correlation functions.
The excitation spectrum of $\widehat{H}^{\ }_{RK}$
depends crucially on whether $\alpha>0$ (locality) or $\alpha=0$
(nonlocality) and can be simulated using classical Monte Carlo methods.
In the ferromagnetic phase, we have argued 
that the relaxation time for a generic state to decay into the ground state 
diverges exponentially with the system size, 
a trademark of quantum glassiness,
when the coupling of $\widehat{H}^{\ }_{RK}$ 
to a bath preserves the condition of locality. 
Thus the system prepared in a state out of equilibrium while in the 
ferromagnetic phase cannot decay into the exact ground state 
$|\Psi^{\ }_{RK}\rangle$ by spontaneous emission of phonons from the bath 
and stalls into a mixed state however small the bath temperature is.

\section*{Acknowledgments}

We would like to thank C.~Henley for his careful reading of this manuscript
and his insightful comments.
This work is supported in part by the NSF Grants DMR-0305482 and DMR-0403997
(C.C.~and C.C.).


\begin{thebibliography}{99}
\bibitem{modulo3vsmodulo6}
The three-coloring constraint on an arbitrary lattice 
with coordination number $3$ translates to the condition 
that the magnetization of any closed path of the lattice 
is an integer multiple of $3$.
(Ref.~\onlinecite{Difrancesco94}) 
This is satisfied if the magnetization of any closed path of minimal 
length is an integer multiple of $3$. 
When the lattice is chosen to be a planar Bravais lattice, i.e., 
the honeycomb lattice, the number of spins along closed paths of 
minimal length paths is even, namely $6$, 
and the condition becomes that the magnetization 
of any elementary hexagon is $0$ or $\pm6$.
\end{thebibliography}
\end{document}